\documentclass[english]{elsarticle}
\usepackage[T1]{fontenc}
\usepackage[latin9]{inputenc}
\usepackage{geometry}
\usepackage{float}
\usepackage{amsmath}
\usepackage{amssymb}
\usepackage{graphicx}
\usepackage{setspace}
\usepackage{esint}
\usepackage[numbers]{natbib}

\makeatletter
\numberwithin{equation}{section}
\numberwithin{figure}{section}

\makeatother

\usepackage{babel}
\begin{document}

\title{A $\text{\ensuremath{\kappa}}$-deformed Model of Growing Complex
Networks with Fitness}

\author{Massimo Stella%
\thanks{Corresponding author's email: massimo.stella@inbox.com.%
}, Markus Brede}

\address{Institute for Complex Systems Simulation, Department of Electronics
and Computer Science, University of Southampton, University Rd, Southampton
SO17 1BJ, United Kingdom.}
\begin{abstract}
The Barab\'{a}si-Bianconi (BB) fitness model can be solved by a mapping
between the original network growth model to an idealized bosonic
gas. The well-known transition to Bose-Einstein condensation in the
latter then corresponds to the emergence of ``super-hubs'' in the
network model. Motivated by the preservation of the scale-free property,
thermodynamic stability and self-duality, we generalize the original
extensive mapping of the BB fitness model by using the nonextensive
Kaniadakis $\kappa$-distribution. Through numerical simulation and
mean-field calculations we show that deviations from extensivity do
not compromise qualitative features of the phase transition. Analysis
of the critical temperature yields a monotonically decreasing dependence
on the nonextensive parameter $\kappa$. 

\end{abstract}
\maketitle

\section*{Introduction}

Over the last twenty years, research about complex networks has yielded
many insights into a large number of real-world systems in various
contexts, with systems as diverse as the World Wide Web, social media,
power grids, transportation networks and gene regulation networks
being prime examples, cf. \citep{Newman2003,Caldarelli2007,Albert:2001:SMC:933363}
for reviews on the field. The network paradigm has proven very useful
in quantifying the topology of interactions in these systems. In recent
years the interest of the scientific community has shifted from analysis
of purely static structures towards attempts at gaining insights into
dynamically evolving or optimized networks \citep{Caldarelli2007}.
In many applications statistical physics has provided a powerful toolbox,
sometimes discovering surprising parallels between networked systems
and other physical systems. One example of such a parallel is the
Barab\'{a}si-Bianconi (BB) model \citep{Bianconi2001Fit,Bianconi2001},
which describes a process of network evolution guided by a combination
of preferential attachment and intrinsic fitness properties of nodes
\citep{Caldarelli2007}. 

In \citep{Bianconi2001}, a mapping between the growing network and
a bosonic gas undergoing a Bose-Einstein condensation was proposed,
realized by means of extensive statistical mechanics, and solved via
a mean field approach. No interactions between particles or energy
level transitions were contemplated. Interesting behaviour is found
when introducing a fictitious temperature parameter $T$ regulating
the network dynamics. Even for $T\rightarrow0$ the bosonic gas counterpart
exhibits a ground state in which only half of the particles reach
the minimum energy level, the others being scattered in fixed positions
throughout the whole energy spectrum \citep{Bianconi2001,Godreche2010}.
This anomalous thermodynamic behavior is incompatible with the Boltzmann
weight used for the original mapping. In fact, a purely physical perspective
would suggest that in equilibrium all the bosons of the Bose-Einstein
condensate populate the minimum energy level in the ground state.
This anomaly has motivated research into applications of deformed
non-Gaussian statistics \citep{Tsallis2009,Clementi11} and it is
of interest to generalize the mapping between bosonic gases and growing
networks by recurring to a nonextensive deformation of the original
equilibrium Boltzmann-Gibbs distribution. Even though Information
Theory seems to suggest a link between non-ergodic behavior and nonextensive
statistics in nature, the emergence of non-Gaussian statistics for
complex systems, as, e.g., growing complex networks, has not yet been
fully understood \citep{Thurner2012}.

To the best of our knowledge one recent study \citep{CinesiSu2011}
of the BB model using Tsallis' $q$-statistics \citep{Tsallis2009}
is the only previous investigation of extensions of the BB model to
nonextensive statistics. Properties of the $q$-statistics make it
very difficult to find exact results and the insights gained by \citep{CinesiSu2011}
are limited to numerical simulation. Further, it has been argued \citep{Kaniadakis2002}
that non-extensive statistics should have the following characteristics:
(i) preservation of scale-free property, (ii) self-duality, and (iii)
thermodynamic stability. Even though the Tsallis $q$-exponential
gives rise to power-laws in many real world modelling applications
\citep{Tsallis2009} it does not satisfy self-duality. This gives
an added interest to the Kanidakis $\kappa$-distribution which meets
all three requirements. The $\kappa$-deformed statistical mechanics
was originally proposed in the context of non-linear kinetics in particle
systems and is deeply linked to the structure of special relativity
\citep{Kaniadakis2002}. In the last decade, $\kappa$-deformed statistics
have been successfully applied to model the distribution of stellar
rotational velocities of dwarf stars \citep{Rotational09}, cosmic
ray fluxes \citep{Kaniadakis2002}, the formation of Quark-Gluon plasmas
\citep{Plasma03}, and the income distribution of the USA, UK, Germany
and Italy \citep{Clementi11}. Moreover, the use of $\kappa$-deformed
statistics has led to some insight \citep{Aliano2002} in addressing
the inadequacy of the Bose-Einstein distribution in predicting the
fluid-superfluid transition temperature in $^{4}He$ . The latter
gives an additional motivation for our application of $\kappa$-deformed
statistics to models of network formation.

In this article we show that the Kaniadakis $\kappa$-distribution
can be used to generalize the BB bosonic mapping. In contrast to previous
numerical results based on $q$-statistics \citep{CinesiSu2011},
analytical mean-field results can be retrieved with the $\kappa$-deformed
distribution. Our findings show that the use of the Kaniadakis $\kappa$-distribution
does not alter the qualitative presence of condensation. We show both
analytically and numerically that the main influence of the nonextensive
parameter is a systematic shift of the critical temperature with $\kappa$.

Our paper is organized as follows. In Section I we briefly review
some of the properties of the BB fitness model and its original extensive
mapping to a Bose-Einstein condensate. In Section II we summarize
some basic results of the Kaniadakis $\kappa$-distribution and generalize
them to the nonextensive mapping. Finally, in Section III numerical
results for the critical temperatures for different values of the
nonextensive parameter are presented and compared to analytical findings.
In the same section, we will also analyze the degree distribution
of our generalized model, by using both analytical and numerical methods.

\section{The Barab\'{a}si-Bianconi Fitness Model}

The BB model \citep{Bianconi2001,Bianconi2001Fit} is a model that
describes network evolution as an incremental growth process. At each
time step a new node, say $j$, is added to the growing network and
it is assigned a fitness $\eta_{j}$, i.e. a positive real number
randomly drawn from a distribution $\rho\left(\eta\right)$. Then
connections from $j$ to $m$ of the $n$ old nodes are formed one at a time,
such that each time old nodes can attract a link from the new node
with a probability given by
\begin{equation}
\Pi\left(k_{i},\eta_{i}\right)=\frac{\eta_{i}k_{i}}{\overset{n}{\underset{j=1}{\sum}}\eta_{j}k_{j}},\label{eq:0}
\end{equation}
which is proportional to both the fitness $\eta_{i}$ and the degree
$k_{i}$ of the node. This preferential attachment dynamics consistently
reproduces some features present in many real world networks \citep{Caldarelli2007,Newman2003},
which exhibit big hubs with different age. The fitness parameter influences
the competition for new connections. Without it, as in the ``pure''
preferential attachement model \citep{Barabsi1999}, ``older'' nodes
are always on average more connected than younger nodes. Including
the fitness aspects, highly connected nodes can be old nodes, but
could also be ``young'' nodes with high intrinsic fitness. One of
the most interesting features of the Barab\'{a}si-Albert (BA) and BB models is the fact
that they both can produce scale-free networks with power-law degree
distributions $P\left(k\right)$ \citep{Bianconi2001,Bianconi2001Fit},
namely $P\left(k\right)\sim Ax^{-\gamma}$ for $x\rightarrow+\infty$,
with scaling exponents $\gamma$ typically between 2 and 3 and compatible
with those of many natural and human systems \citep{Caldarelli2007,Newman2003}.

\subsection{Extensive mapping with a Bose-Einstein system and the Network Condensation}

In \citep{Bianconi2001}, Bianconi realized a mapping between the
asymptotic structure of the network and an hypothetical Bose gas by
setting $\eta_{i}=e^{-\beta\epsilon_{i}}$. In this formulation $\beta$
is the inverse of a fictitious network temperature $T$ and $\epsilon_{i}$
is the energy level for node $i$. Every node of the network represents
one energy level while every edge between two nodes $i$ and $j$
represents two non-interacting particles, one on level $\epsilon_{i}$
and the other on level $\epsilon_{j}$. It can be shown \citep{Bianconi2001,Bianconi2001Fit,Godreche2010,Ferretti2008}
that preferential attachment driven by the probability $\Pi$ undergoes
a phase transition at some critical temperature $T_{c}$, which is
formally identical to that of Bose-Einstein condensation. At high
temperatures, $T>T_{c}$, even in the thermodynamic limit the hubs
of the network participate in a competition for the ``\emph{survival-of-the-fittest}''
while at lower temperatures, for $T<T_{c}$, a unique ``super-hub''
eventually emerges and wins the competition \citep{Ferretti2008}
(this represents a ``\emph{winner-takes-all}'' phenomenon encountered
in many real-world applications \citep{Caldarelli2007}). By using
the network mapping of edges to a physical system of $2N$ bosons,
it is evident that eventually only half of them can ``reach'' the
fundamental energy level of the network, which corresponds to the
highest fitness present.

\section{Kaniadakis Mapping }

The Kaniadakis $\kappa$-distribution \citep{Kaniadakis2002}, was
originally introduced in 2002 and later re-obtained within the framework
of a Jaynes Maximum Entropy principle \citep{Kaniadakis2009}, which
starts from a generalized system entropy $S_{\kappa}$ given by
\begin{equation}
S_{\kappa}=-\underset{l}{\sum}p_{l}\ln_{\kappa}\, p_{l}=-\underset{l}{\sum}p_{l}\frac{\left(p_{l}\right)^{\kappa}-\left(p_{l}\right)^{-\kappa}}{2\kappa}\label{eq:2}
\end{equation}
with $p_{l}$ being the probability of the system being in state $l$
and $\kappa\in\left(-1,1\right)$. It can be shown that the generalized
logarithm $\ln_{\kappa}\left(x\right)$ is self-scaling and self-dual
\citep{Kaniadakis2009}, as the ordinary logarithm $\ln\left(x\right)$
of the Boltzmann-Shannon entropy $S=-\sum p_{l}\ln\left(p_{l}\right)$. 

The parameter $\kappa$ can be seen as a deviation from extensivity,
as $\ln_{\kappa}\left(x\right)$ recovers the extensive case when
$\kappa\rightarrow0$. The inverse of the generalized logarithm is
the generalized exponential \citep{Kaniadakis2002,Kaniadakis2009},
\begin{equation}
e_{\kappa}^{x}=\left(\kappa x+\sqrt{1+\kappa^{2}x^{2}}\right)^{\frac{1}{\kappa}}
\end{equation}
which has power-law tails given by the asymptotics $e_{\kappa}^{x}\sim2\left|\kappa x\right|^{\frac{\pm1}{\left|\kappa\right|}}$
for $x\rightarrow\pm\infty$ \citep{Kaniadakis2009}. It has been
shown \citep{Kaniadakis2002} that this $\kappa$-deformed exponential
is a positive monotonically increasing function which is symmetric
with respect to the nonextensive parameter $\kappa$ ($e_{\kappa}^{x}=e_{-\kappa}^{x}$)
and  satisfies $\forall x\in\mathbb{R},\; e_{\kappa}^{x}e_{\kappa}^{-x}=1$
and $\forall x,a\in\mathbb{R},\; e_{\kappa}^{ax}=\left(e_{a\kappa}^{x}\right)^{a}$.

Mimicking the mapping of the original BA model, we consider the fitness
$\eta_{i}$ as
\begin{equation}
\eta_{i}=e_{\kappa}^{-\beta\epsilon_{i}}=\left[\kappa\left(-\beta\epsilon_{i}\right)+\sqrt{1+\kappa^{2}\beta^{2}\epsilon_{i}^{2}}^{\frac{1}{\kappa}}\right]\label{eq:fitndef}
\end{equation}
where the energy level $\epsilon_{i}=-\beta^{-1}\ln_{\kappa}\eta_{i}$
can be thought of as a transformation of a random variable distributed
according to the distribution $g\left(\epsilon\right)$, the latter
having the physical interpretation of an energy level density. In
order for the network to undergo a Bose-Einstein phase transition,
it is required that $g\left(\epsilon\right)\rightarrow0$ when $\epsilon\rightarrow0$
\citep{Bianconi2001,Bianconi2001Fit,CinesiSu2011}. With this generalization,
Eq. (\ref{eq:0}) for the degree $k_{i}$ becomes 
\begin{equation}
\frac{dk_{i}}{dt}=m\frac{e_{\kappa}^{-\beta\epsilon_{i}}k_{i}\left(\epsilon_{i},t_{i},t\right)}{\underset{j}{\sum}e_{\kappa}^{-\beta\epsilon_{j}}k_{j}\left(\epsilon_{j},t_{j},t\right)}=m\frac{e_{\kappa}^{-\beta\epsilon_{i}}k_{i}\left(\epsilon_{i},t_{i},t\right)}{Z_{\kappa}\left(t\right)}\label{eq:1}
\end{equation}
where $Z_{\kappa}$ is our network partition function. We want to
look for solutions $k_{i}$ which mimic the analytical form of the
solution for the BA model \citep{Bianconi2001},
\begin{equation}
k_{i}\left(\epsilon_{i},t_{i},t\right)=A_{m}\left(\frac{t}{t_{i}}\right)^{f_{\kappa}\left(\epsilon_{i}\right)}\label{eq:dynexp}
\end{equation}
where $A_{m}$ depends on the parameter $m$ while $f_{\kappa}=f_{\kappa}\left(\epsilon_{i}\right)$
is some still unknown dynamic exponent strictly bounded by 0 and 1.
This functional form implies that the number of connections of every
node increases in time ($f_{\kappa}>0$) but not as fast or faster
than $t$ itself ($f_{\kappa}<1$). We can solve the above differential
equation by applying a mean field approximation, in which we replace
the partition sum $Z_{\kappa}$ by its average $\left\langle Z_{\kappa}\right\rangle $,
\begin{equation}
\left\langle Z_{\kappa}\right\rangle =A_{m}\int d\epsilon g\left(\epsilon\right)\frac{e_{\kappa}^{-\beta\epsilon}}{1-f_{\kappa}\left(\epsilon\right)}\left(t-t^{f_{\kappa}\left(\epsilon\right)}\right)\;.\label{eq:3}
\end{equation}

For this purpose we introduce a chemical potential $\mu$ and a mean
fugacity $z_{\kappa}=e_{\kappa}^{\beta\mu}$ and hence by using the
thermodynamical stability of the deformed $\kappa$-exponential \citep{Kaniadakis2009}
one obtains: 
\begin{equation}
\left\langle Z_{\kappa}\right\rangle \sim A_{m}\frac{t}{z_{\kappa}}+O\left(t^{\alpha}\right)\;.\label{eq:fuganum}
\end{equation}

Here $\alpha=f_{\kappa}\left(\epsilon\right)-1<0$ and then $t^{\alpha}\rightarrow0$
in the thermodynamic limit. Comparing to the extensive case (obtained
when $\kappa\rightarrow0$), it has to be $\left\langle Z_{\kappa}\right\rangle \overset{\kappa\rightarrow0}{\longrightarrow}mt/z$
for big enough networks. As by construction $A_{m}$ is independent
of $\kappa$ we can choose $A_{m}=m$. Within this approximation Eq.
(\ref{eq:1}) has a solution of the same form as given in Eq. (\ref{eq:dynexp})
with a dynamic exponent $f_{\kappa}\left(\epsilon_{i}\right)=e_{\kappa}^{\beta\mu_{\kappa}}e_{\kappa}^{-\beta\epsilon_{i}}$.
As $f_{\kappa}\left(\epsilon_{i}\right)\overset{\kappa\rightarrow0}{\longrightarrow}e^{-\beta\left(\epsilon_{i}-\mu\right)}$
these results recover what is already known in the extensive limit
\citep{Bianconi2001}.

By the definition of $z_{k}$ and exploiting the self-duality of the
$\kappa$-exponential \citep{Kaniadakis2009}, we have that
\begin{equation}
I_{\beta}\left(\mu_{\kappa}\right)=\int d\epsilon g\left(\epsilon\right)\frac{1}{e_{\kappa}^{-\beta\mu_{\kappa}}e_{\kappa}^{\beta\epsilon}-1}=1\:.
\end{equation}

In the integrand of this integral we can recognize the distribution
$n_{\kappa}\left(\epsilon\right)$, 
\begin{equation}
n_{\kappa}\left(\epsilon\right)=\frac{1}{e_{\kappa}^{-\beta\mu_{\kappa}}e_{\kappa}^{\beta\epsilon}-1}\underset{\kappa\rightarrow0}{\longrightarrow}n\left(\epsilon\right)=\frac{1}{e^{\beta\left(\epsilon-\mu\right)}-1}
\end{equation}
which, in our context, represents a deformed Bose-Einstein distribution,
with the very same properties of monotonicity as the original Bose-Einstein
distribution $n\left(\epsilon\right)$ in the non-extensive case. 

Contrary to the case of the fitness model generalized with the Tsallis
entropy \citep{CinesiSu2011}, the Bose-Einstein distribution $n_{\kappa}\left(\epsilon\right)$
can be rewritten in a more insightful way by using properties of the
$\kappa$-deformed exponential \citep{Kaniadakis2002,Kaniadakis2009}.
In fact, by using $e_{\kappa}^{x}=\exp\left(\frac{1}{\kappa}\text{arcsinh}\,\kappa x\right)$,
one can define a $\kappa$-deformed sum $\overset{\kappa}{\oplus}$
\citep{Kaniadakis2002}, such that $\left(\mathbb{R},\overset{\kappa}{\oplus}\right)$
is an Abelian group and
\begin{equation}
e_{\kappa}^{x}e_{\kappa}^{y}=e_{\kappa}^{\left(x\overset{\kappa}{\oplus}y\right)}=\exp\left(x\sqrt{1+\kappa^{2}y^{2}}+y\sqrt{1+\kappa^{2}x^{2}}\right)\;.
\end{equation}

This recovers the usual property of exponentials in the extensive
limit, i.e. $e_{\kappa}^{x}e_{\kappa}^{y}\overset{\kappa\rightarrow0}{\longrightarrow}e^{x+y}$.
From a physical point of view, it has to be remarked that this $\kappa$-deformed
sum emerges naturally from the composition law of relativistic momenta
in the framework of special relativity \citep{Kaniadakis2002}. It
is straightforward to generalize these formulas to a $\kappa$-deformed
subtraction $\overset{\kappa}{\ominus}$ such that, $x\overset{\kappa}{\ominus}y=x\overset{\kappa}{\oplus}(-y)$
and 
\begin{equation}
n_{\kappa}\left(\epsilon\right)=\frac{1}{e_{\kappa}^{\beta\left(\epsilon\overset{\kappa}{\ominus}\mu\right)}-1}
\end{equation}
is formally equivalent to a Bose-Einstein-like distribution for a
system whose energy levels are distributed according to a certain
density $g\left(\epsilon\right)$. Here, however, the differences
between the single particle energy and the chemical potential of the
whole ensemble have to be interpreted in a $\kappa$-deformed way. 

Exactly as in the original Bose-Einstein distribution, because of
the physical meaning of $n_{\kappa}\left(\epsilon\right)$ as the
probability of having a particle with energy between $\epsilon$ and
$\epsilon+d\epsilon$, it has to be
\begin{equation}
n_{\kappa}\left(\epsilon\right)\geq0\rightarrow\epsilon\overset{\kappa}{\ominus}\mu\geq0\rightarrow\frac{\epsilon}{\sqrt{1+\kappa^{2}\epsilon^{2}}}\geq\frac{\mu}{\sqrt{1+\kappa^{2}\mu^{2}}}
\end{equation}
which, as in the extensive case, implies that $\mu\leq\epsilon_{min}=0$
for bosonic systems.

Similar to the originally proposed model in this generalized BB model
a Bose-Einstein phase transition occurs as long as $g\left(\epsilon\right)\rightarrow0$
for $\epsilon\rightarrow0$, i.e. if 
\begin{equation}
I_{\beta}\left(\mu=0\right)=\int d\epsilon g\left(\epsilon\right)n_{\kappa}\left(\epsilon\right)<1\;.
\end{equation}

We will present analytical and numerical estimates for the critical
temperature in the next section.

\section{Generalized Barab\'{a}si-Bianconi Model: Numerical and Analytical Computations}

\subsection{Critical Temperature and Nonextensivity Parameter}

Following arguments of \citep{Bianconi2001} it can be seen that the
Bose-Einstein condensation on the network corresponds to an almost
zero value for the chemical potential $\mu$, which can be computed
using a preliminary result we obtained while solving Eq. (\ref{eq:1}),
i.e. 
\[
\mu\sim-\frac{1}{\beta}\text{ln}_{\kappa}\left(\frac{\left\langle Z_{\kappa}\right\rangle }{mt}\right)
\]
where the $\kappa$-deformed logarithm was defined as in Eq. (\ref{eq:2})
and where the last approximation was made in order to solve Eq. (\ref{eq:1}).
Estimates for the average partition function $\left\langle Z_{\kappa}\right\rangle $
over an ensembles of moderately sized networks can be obtained numerically.

We choose a simple distribution for $g\left(\epsilon\right)$ which
is compatible with the occurrence of condensation and which also allows
for easy comparisons with the numerical results of \citep{Bianconi2001,Godreche2010},
i.e. we use $g\left(\epsilon\right)=2\epsilon$ with corresponding
energies $\epsilon\in(0,1)$. To obtain the phase diagram $\left|\mu\right|$
vs $T$ we carried out numerical simulations for varying network size
$N$. As stated in \citep{Bianconi2001}, and confirmed in \citep{CinesiSu2011},
for temperatures higher than the critical temperature $T_{c}^{\left(\kappa\right)}$
(in which $\mu$ vanishes) the network is in the fit-get-rich (FGR)
phase, while under $T_{c}^{\left(\kappa\right)}$ one or more giant
hubs appear and the nodes ``condensate'' their links to them, the
whole network being in the winner-takes-all phase. 

\begin{figure}[H]
\begin{centering}
\includegraphics[width=12cm]{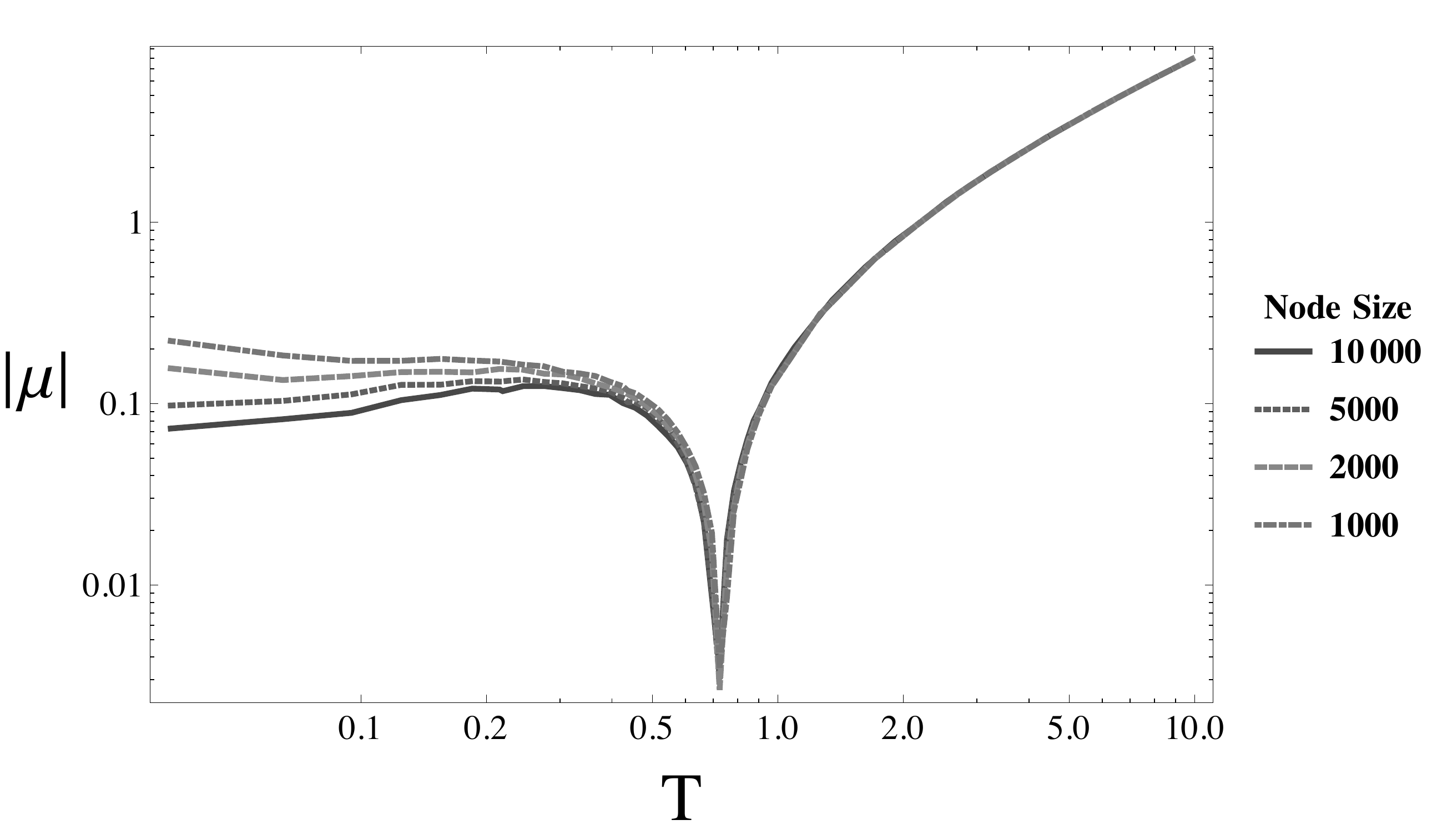}
\par\end{centering}

\caption{Phase diagram for different size ensembles of the generalized BB network,
with $N=\{10^{3},2\cdot10^{3},5\cdot10^{3},10^{4}\}$ nodes and respectively
averaged over $\{800,400,200,120\}$ different configurations, all
with $m=2$ and $\kappa=0.9$. The critical temperature $T_{c}^{\left(0.9\right)}\simeq0.76$
is statistically lower than the one reported in the extensive case
\citep{Bianconi2001}, for $\kappa=0$, $T_{c}\simeq0.8$. A scaling
phenomenon is evident, following the decrease in temperature.}

\end{figure}

By averaging over 800 network realizations we determined a critical
temperature $T_{c}^{\left(0.9\right)}=0.732\pm0.002$ for a network
ensemble with $m=2$, $\kappa=0.9$ and $N=1000$. This temperature
is considerably lower than the one observed in the extensive ($\kappa=0$)
case, in which $T_{c}^{\left(0\right)}=0.792\pm0.002$, a value compatible
with the previous finding of $T_{c}^{+}=0.8\pm0.1$ in \citep{Bianconi2001},
and in \citep{CinesiSu2011}. 

\begin{figure}[H]
\begin{centering}
\includegraphics[width=12cm]{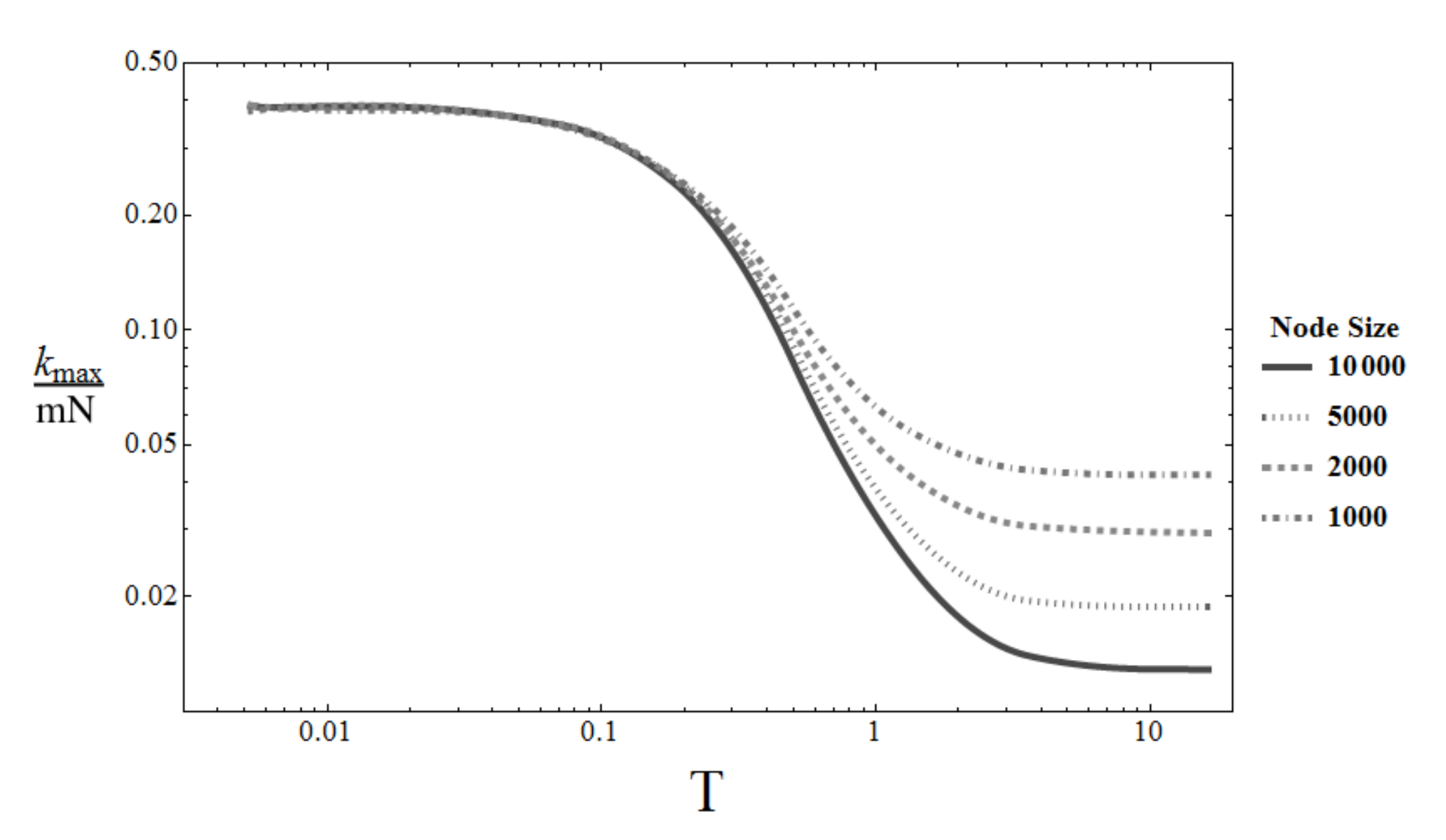}
\par\end{centering}

\caption{Average fraction of connections for the biggest hub, $k_{max}/mt$,
inside a generalized BB network of $N=\{10^{3},2\cdot10^{3},5\cdot10^{3},10^{4}\}$
nodes, respectively averaged over $\{1200,1200,600,150\}$ configurations,
all with $m=2$ and $\kappa=0.9$. The average fraction of connections
$k_{max}/mt$ does not vanish, in the ``winner-takes-all'' condensate
phase, and the system undergoes a gelation phenomenon, in which a
large fraction of the nodes inside the network are connected to the
same ``gel'' hub \citep{Krapiv00}. As in the phase diagram, a scaling
phenomenon is evident in the condensate phase.\label{fig:Average-fraction-of}}
\end{figure}

\begin{figure}[H]
\begin{centering}
\includegraphics[width=9cm]{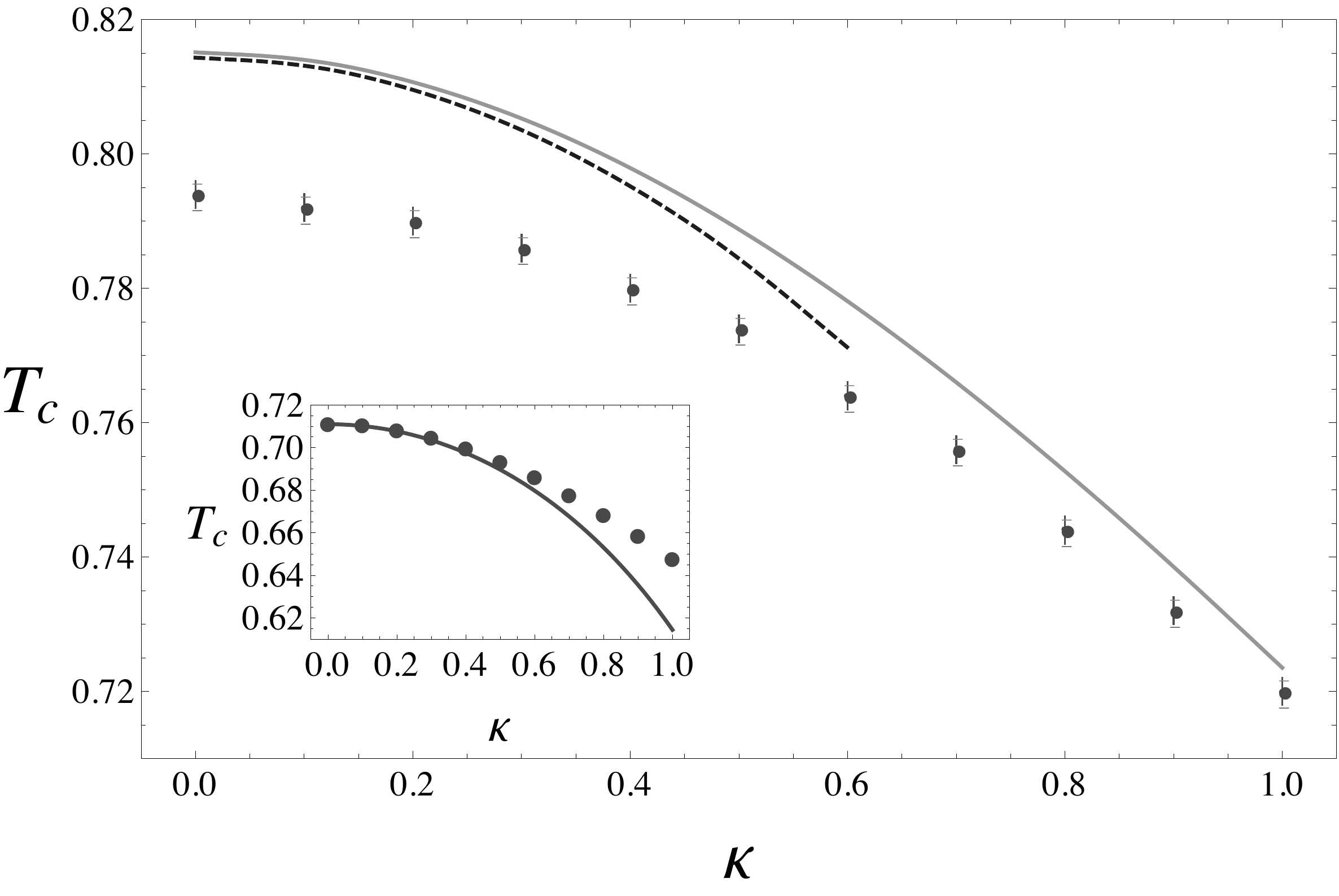}
\par\end{centering}

\caption{Main Figure: Critical temperature $T_{c}^{\kappa}$ vs nonextensive
parameter $\kappa$ (points). The numerical results are based on an
ensemble of 800 networks with $N=10^{3}$ nodes. The temperature for
$\kappa=0.9$ is $(0.732\pm0.002)$ and it is statistically compatible
with the previous result $T_{c}^{\left(0.9\right)}=(0.73\pm0.03)$
for an ensemble of 90 networks with $N=10000$. The upper bound $T_{c\kappa}^{*}+\Delta T$
is computed numerically (continuous line) and by using our analytical
upper bound (dashed line). Inset: Numerically computed $T_{c\kappa}^{*}$
(dots) vs our analytical upper bound (continuous line). Because of
the truncated McLaurin expansion, our approximation underestimates
the critical temperature and breaks around $\kappa\sim0.5$. \label{fig:Critical-temperature-}}
\end{figure}

The simulation data plotted in Figure \ref{fig:Average-fraction-of}
show that the phase transition is present in the generalized model,
as predicted by our theoretical findings. However it is also evident
that the critical temperature of Bose-Einstein condensation depends
on the $\kappa$ parameter. To investigate the dependence of $T_{c}$
on $\kappa$ we obtained numerical estimates which are shown in Figure
\ref{fig:Critical-temperature-}, suggesting that $T_{c}$ is a monotonically
decreasing function of $\kappa$. 

We next derive an analytical estimate for the dependence of $T_{c}$
on $\kappa$. This can be done based on the above mean field arguments,
by noting the condition for the phase transition at the critical temperature
in the thermodynamical limit $T_{c\kappa}^{*}=1/\beta_{c\kappa}$,
in which $\mu$ has its maximal value, i.e.

\begin{equation}
I_{\beta_{c\kappa}}\left(\mu=0\right)=2\intop_{\epsilon_{min}(t)}^{\epsilon_{max}(t)}d\epsilon\frac{\epsilon}{e_{\kappa}^{\beta_{c\kappa}\epsilon}-1}\leq1\label{eq:tn}
\end{equation}
where $\epsilon_{min}(t)$ and $\epsilon_{max}(t)$ are the minimum
and the maximum energy levels (or the maximum and the minimum fitnesses)
at time $t$. Unfortunately, different to the $\kappa=0$ case, the
approximation of extending the integral from 0 to infinity is not
valid in our case. This is due to the asymptotics of the $\kappa$-deformed
exponential which vanishes only as $e_{\kappa}^{x}\sim2\left|\kappa x\right|^{1/\left|k\right|}$
and not exponentially when $x\rightarrow+\infty$. 

However, approximating $\epsilon_{min}(t)\simeq0$ and $\epsilon_{max}(t)\simeq1$
and performing a McLaurin expansion of the integrand function up to
the second order leads to:

\begin{equation}
\frac{3}{8}-\frac{\sqrt{65-32\kappa^{2}}}{24}<T_{c\kappa}^{*}<\frac{3}{8}+\frac{\sqrt{65-32\kappa^{2}}}{24}\;,\:\kappa\in[0,1]\;.
\end{equation}

Notice that this analytical approximation is an underestimation of
the lower and upper bounds on the critical temperature in the thermodynamic
limit. In particular for $\kappa=0$ our approximation retrieves a
rounded upper bound for the critical temperature $T_{c\kappa}^{*}\leq0.711$,
which actually reproduces the first three digits of the approximation
for the effective temperature $T_{c}^{\star}\simeq0.711$ computed
in \citep{Godreche2010}, by both means of analytic results and numerical
experiments.

In order to compare our analytical findings with the numerical ones,
we have to be aware of the finite size effects $\Delta T_{c}^{\kappa}$
\footnote{In our computations, this correction was calculated by the means of
numerical integration and by using a $T_{c\kappa}^{*}$ numerically
retrieved from Eq. (\ref{eq:tn}). We would also like to point out
an error in the original paper from \citep{Godreche2010}, where $\mu(T)$
in Eq. (2.20) should clearly have an exponential in the numerator
of the integrand function.%
} in the numerical estimates for the critical temperatures $T_{c}^{\kappa}\simeq T_{c\kappa}^{*}+\Delta T$.
Following the results in \citep{Godreche2010}, the finite size correction
$\Delta T_{c}^{\kappa}=\Delta T(N,T_{c\kappa}^{*})$ can be estimated
with a first order series expansion from $I(\beta,\mu)$ at the critical
point. One obtains:
\[
\Delta T\simeq\frac{(T_{c\kappa}^{*})^{4}}{\ln N}\intop_{0}^{1}\frac{\epsilon^{2}e_{\kappa}^{\epsilon/T_{c\kappa}^{*}}}{(e_{\kappa}^{\epsilon/T_{c\kappa}^{*}}-1)^{2}}d\epsilon\;.
\]

We notice that the analytical upper bound is in good agreement with
numerical simulations for low values of $\kappa$. In agreement with
the simulations we thus find that the critical temperature for the
Bose-Einstein condensation decreases with $\kappa$. This is also
compatible with a similar result derived in \citep{Aliano2002}, under
slightly less general assumptions on the functional form of the $\kappa$-deformed
Bose-Einstein distribution.

All in all, it is interesting to note that the deformations inside
the distribution $n_{\kappa}\left(\epsilon\right)$ do not influence
the qualitative presence of the condensation, which is ultimately
rooted in the properties of the energy density $g\left(\epsilon\right)$.
These results suggest that the use of a nonextensive statistics, such
as the Kaniadakis one, does not cause any qualitative change of the
network dynamics. The main influence of a change from extensive to
nonexstensive statistics are quantitative variations in the critical
temperature. This observation is in agreement with the results of
the numerical experiments reported in \citep{CinesiSu2011} for the
Tsallis $q$-distribution. In order to further investigate this claim,
we will proceed with the analysis of the degree distributions of our
generalized fitness model in the next section.

\subsection{Degree Distribution and Nonextensivity Parameter}

In \citep{Bianconi2001Fit}, the authors analyzed the degree distribution
$P(k)$ of a fitness model with an exponential fitness distribution
$\rho(\eta)$, obtaining the following estimate

\begin{equation}
P(k)\propto\int d\eta\rho(\eta)\frac{1}{F(\eta)}\left(\frac{m}{k}\right)^{F(\eta)^{-1}+1}
\end{equation}
where $F(\eta)$ is the dynamical exponent which determines the growth
of node degrees conditional on fitness. Using Eq. (\ref{eq:dynexp})
$F(\eta)$ is to be identified with $f_{\kappa}\left(\epsilon\right)=e_{\kappa}^{\beta(\mu_{\kappa}\overset{\kappa}{\ominus}\epsilon)}$
in our generalized model. According to our definition of fitness $\eta$,
given in Eq. (\ref{eq:fitndef}), we have $e_{\kappa}^{-\beta\epsilon}=\eta$
and $e_{\kappa}^{\beta\mu}=z_{k}$ is the fugacity. Furthermore, $\rho(\eta)=\frac{1}{\beta}e_{\kappa}^{\beta\epsilon}g(\epsilon)$
is the distribution of fitnesses expressed in terms of the energy
level distribution $g(\epsilon)$. These substitutions lead to:

\begin{equation}
P_{\kappa}(k)\propto\intop_{e_{\kappa}^{-\beta}}^{1}d\eta\left(-\frac{2}{\beta^{2}}\frac{\ln_{\kappa}\eta}{\eta}\right)\frac{1}{z_{\kappa}\eta}\left(\frac{m}{k}\right)^{1+\left(z_{\kappa}\eta\right)^{-1}}=\label{eq:5}
\end{equation}
\[
=\intop_{e_{\kappa}^{-\beta}}^{1}d\eta\left(-\frac{2}{\beta^{2}}\frac{\ln_{\kappa}\eta}{\eta}\right)\varphi_{\kappa,k}(\eta)=\mathbb{E}[\varphi_{\kappa,k}(\eta)]\;.
\]

Even if the fictitious temperature in $\beta$ only plays the role
of a control parameter of the model, it actually has some influence
on the degree distribution, entering implicitly in the determination
of the scaling exponent via $z_{\kappa}$ and also via the lower bound
of fitness range. 

$\varphi_{\kappa,k}(x)$ is a convex function for $\eta\in[e_{\kappa}^{-\beta},1]$
when $k>16$ and making use of Jensen's inequality one finds

\begin{equation}
\mathbb{E}[\varphi_{\kappa,k}(\eta)]\geq\varphi_{\kappa,k}\left(\mathbb{E}[\eta]\right)=\frac{1}{z_{\kappa}\mathbb{E}[\eta]}\left(\frac{m}{k}\right)^{1+\left(z_{\kappa}\mathbb{E}[\eta]\right)^{-1}}\;.\label{eq:4b}
\end{equation}

By using the definition of the $\kappa$-deformed logarithm given
in Eq. (\ref{eq:2}) one can now compute the expected value of the
fitness $\mathbb{E}[\eta]$ which is a monotonically increasing function
of the temperature for every $\kappa\in(0,1]$: 

\begin{equation}
\mathbb{E}[\eta]=\frac{1}{\kappa\beta^{2}}\left[\frac{1}{\kappa+1}(e_{\frac{\kappa}{\kappa+1}}^{-\beta(\kappa+1)}-1)+\frac{1}{\kappa-1}(e_{\frac{\kappa}{1-\kappa}}^{-\beta(1-\kappa)}-1)\right]\;.
\end{equation}

The arguments above motivate our estimate for the degree distribution
of a generalized fitness network with a logarithmic distribution of
fitnesses over a finite range, in terms of a lower bound given by
a power law:

\begin{equation}
P_{\kappa}(k)\sim\frac{1}{z_{\kappa}\mathbb{E}[\eta]}\left(\frac{m}{k}\right)^{1+\left(z_{\kappa}\mathbb{E}[\eta]\right)^{-1}}\label{eq:4}
\end{equation}
with the exponent $\gamma_{\kappa,T}=1+\left(z_{\kappa}\mathbb{E}[\eta]\right)^{-1}$
implicitly dependent on network temperature and on the nonextensive
parameter $\kappa$. The fugacity $z_{\kappa}$ can be determined
using Eq. (\ref{eq:fuganum}). 

\begin{figure}[h]
\begin{centering}
\includegraphics[width=9cm]{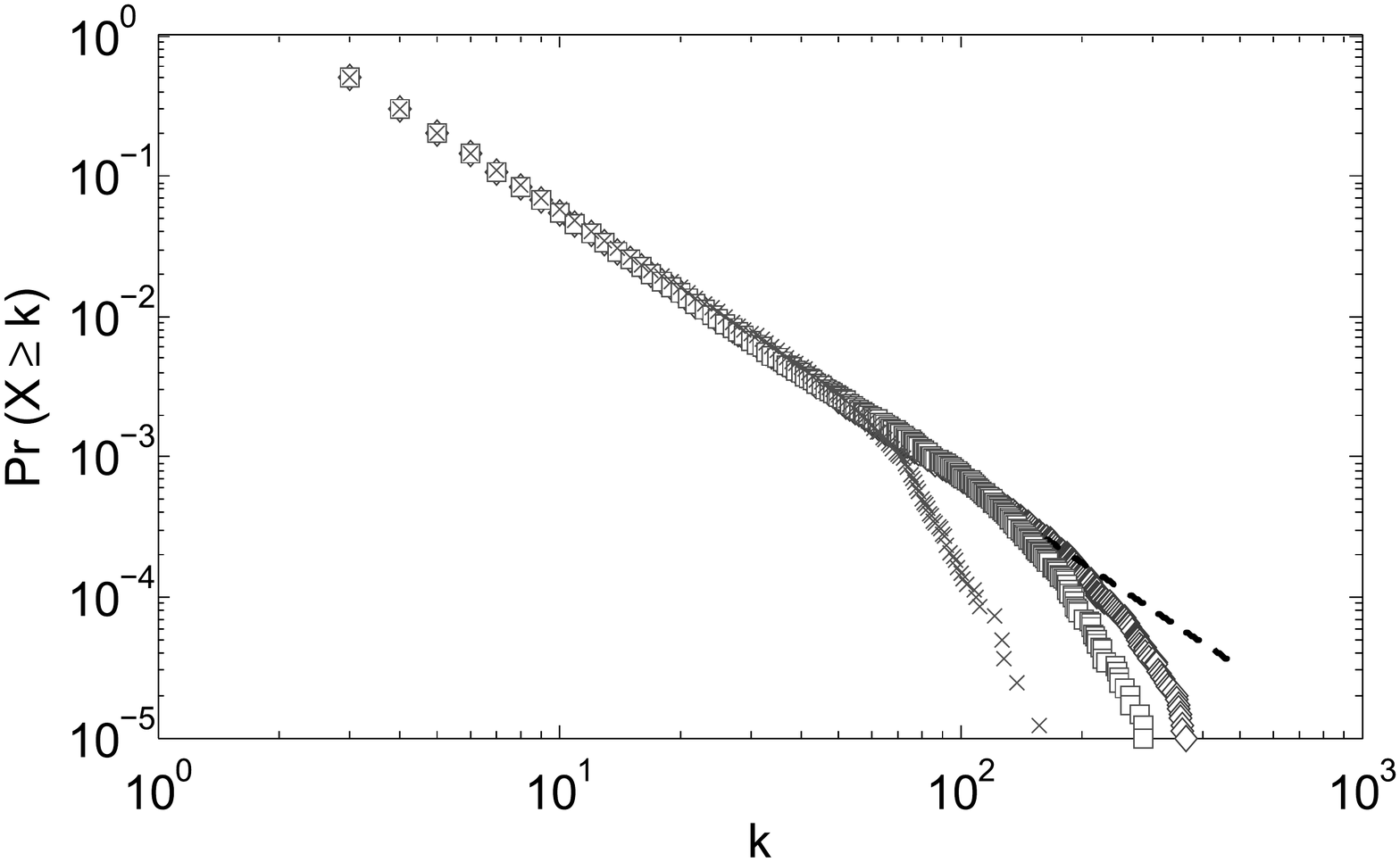}\\
\includegraphics[width=4.5cm]{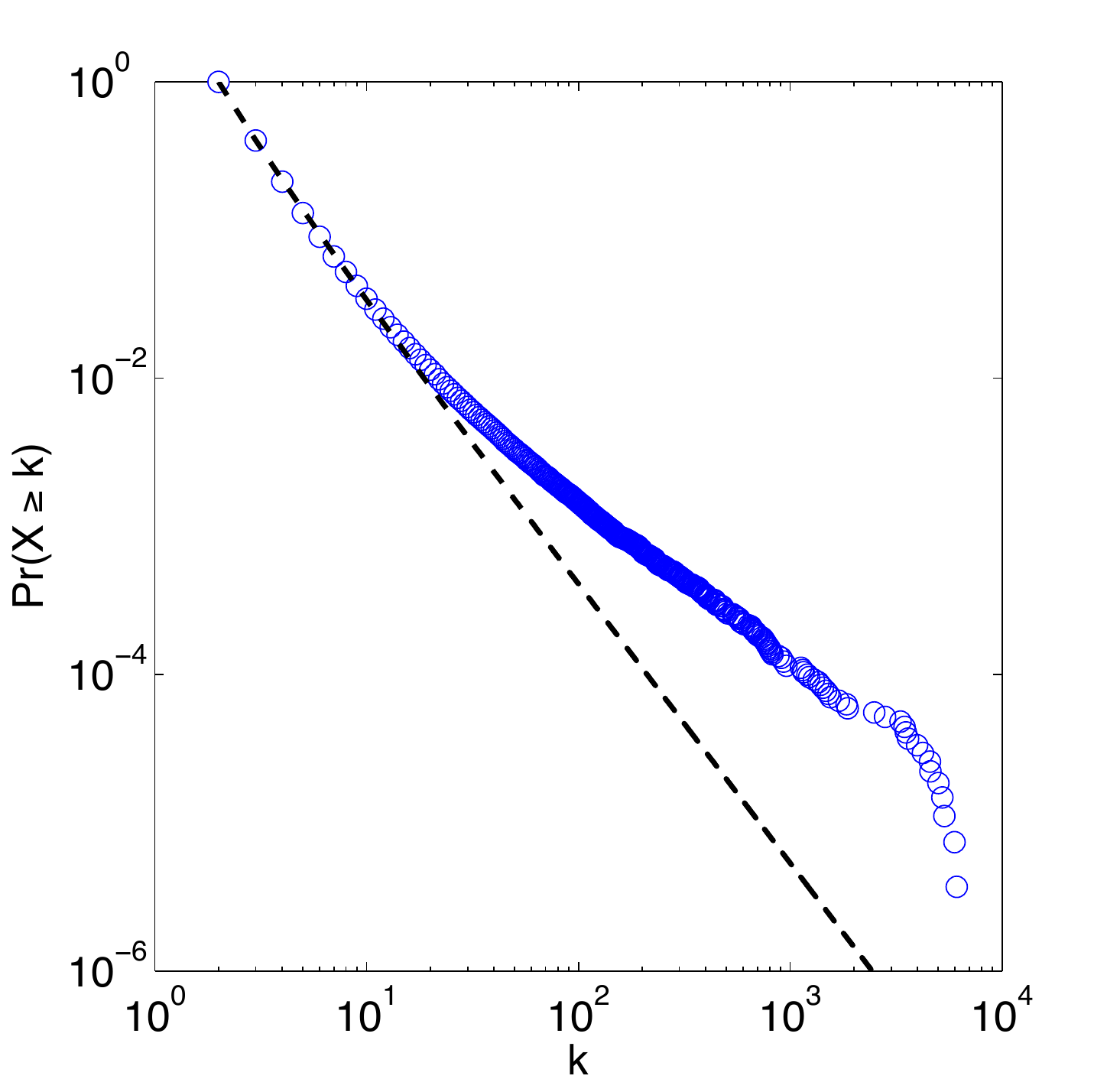}\includegraphics[width=4.5cm]{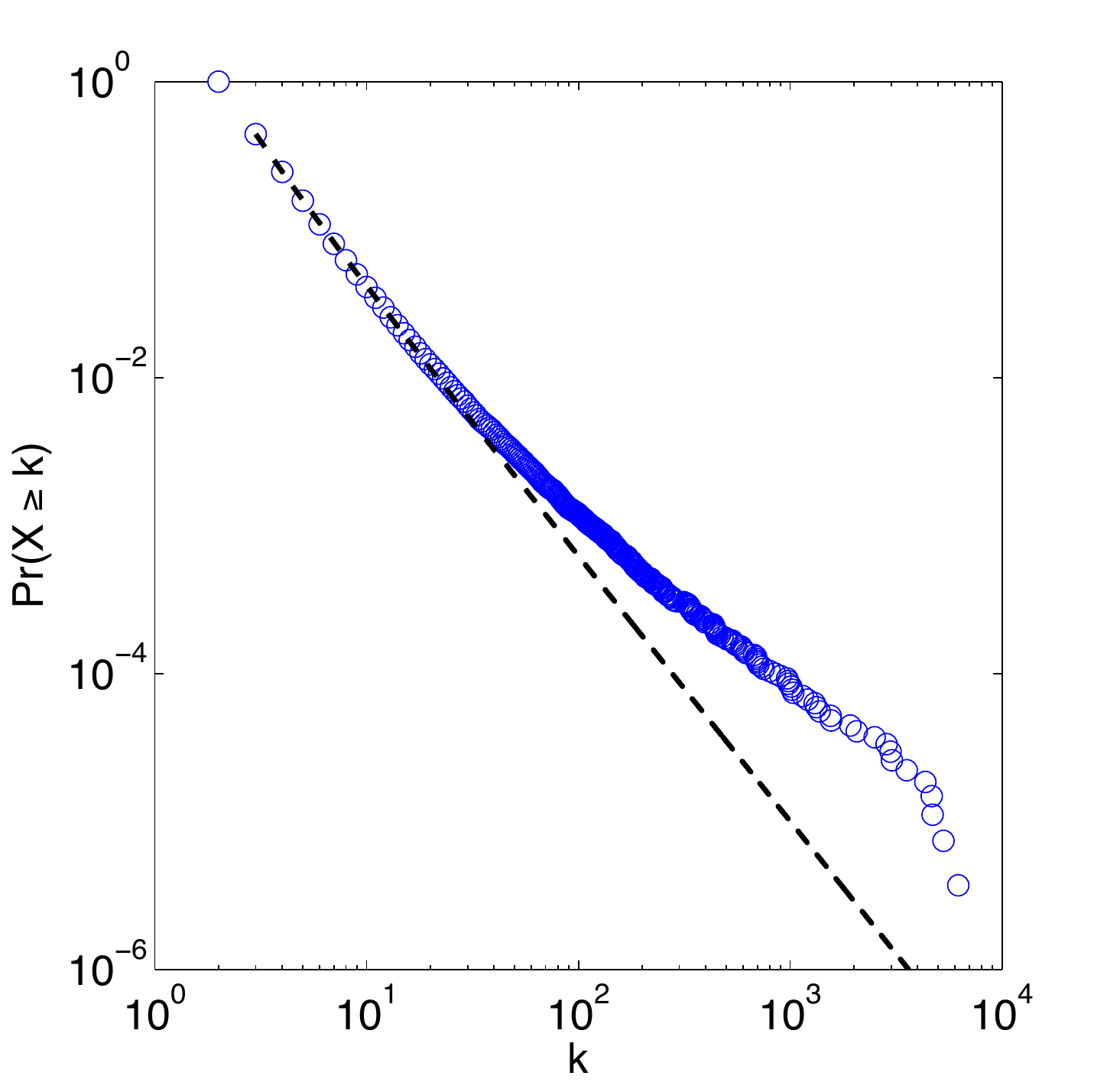}
\par\end{centering}

\caption{Complementary cumulative distribution functions $P(x\geq k)$ for
the degree distribution in a network with $N$ nodes for different
values of temperature and nonextensive parameter. Upper Centre:$\kappa=0.9$
and $T=5$ for $N=10^{3}$ (crosses), $N=5\cdot10^{3}$ (squares)
and $N=10^{4}$. The dashed straight line is a power law with exponent
$\gamma=3$, for reference. The data shows a power-law like behavior
with a cut-off depending on the system finite size. Lower Left: $\kappa=0.2$
and $T=0.5$. Lower Right: $\kappa=0.9$ and $T=0.5$.\label{fig:Left:-Complementary-cumulative}
The dashed line is the estimate based on a power-law assumption, with
scaling exponents respectively $\gamma_{0.2,0.5}\simeq2.86$ and $\gamma_{0.9,0.5}\simeq2.79$.
Notice that the first part of the plot is robust to temperature changes,
as confirmed by the retrieved scaling exponents. Lowering the temperature
stretches the distribution only in the higher degree zone. }
\end{figure}

Numerical simulations of degree distributions show a good level of
agreement with the power-law estimate. In the ``fit-get-rich'' phase
($T=5>T_{c}^{\left(\kappa\right)}\,\forall\kappa\in[0,1]$) three
ensembles of 100 networks of sizes $N=\{10^{3},5\cdot10^{3},10^{4}\}$
clearly display power-law like behavior with a finite-size cut off.
Similar results were cited for upper bounded fitness distributions
in \citep{Bianconi2001Fit}.

In Figure \ref{fig:Left:-Complementary-cumulative} a maximum likelihood
fitting procedure \citep{Clauset} reveals scaling exponents for the
power-law like parts of the plots statistically compatible with the
value $\gamma_{0.2,5}=2.89\pm0.05$ for the ensemble with $N=10^{4}$
and $\kappa=0.9$. Other scaling exponents for lower values of $\kappa$
were retrieved, all close to the exponent ($\simeq3$) estimated from
our analytical upper bound in Eq. (\ref{eq:4}). The maximum-likelihood
procedure we used avoids biases generated by adopting a linear binning
scheme, which could potentially alter or masque the stretched-exponential
region. Our estimate for the scaling parameter is suprisingly close
to the one for a growing network with preferential attachment \citep{Barabsi1999,Albert:2001:SMC:933363},
which is recovered by the generalized model either with the choice
$\rho(\eta)=\delta(\eta-\eta_{0})$ or in the limit $T\rightarrow+\infty$
(in both cases, every node has the same fitness). Furthermore, in
the limit $T\rightarrow\infty$ the $\kappa$-deformed exponential
for the fitnesses recovers the Boltzmann weight, in which case it
can be shown analytically that $1/z_{\kappa}\rightarrow1/z\rightarrow2$
while in the same limit $T\rightarrow\infty$ the average fitness
for our model converges to 1, so that the scaling parameter recovers
our numerical estimate 
\begin{equation}
\gamma_{\kappa,T}\overset{T\rightarrow\infty}{\rightarrow}3\,\forall\kappa\in[0,1].
\end{equation}

Nonetheless, in \citep{Bianconi2001Fit}, it was shown that the fitness
model is not stable under changes of the functional form of the fitness
distribution function: The model can give rise to power laws (uniform
$\rho(\eta)$) or non power-law degree distributions (unbounded exponential
distributions $\rho(\eta)$). 

In the high temperature region $T\gg T_{c}^{\left(\kappa\right)}$,
numerical experiments for the fugacity $z_{\kappa}$, performed on
an ensenmble of 800 networks with $N=10^{3}$, suggest that our estimate
$\gamma_{\kappa,T}$ is effectively independent of both the temperature
and the nonextensive parameter, with $\gamma_{\kappa,T}=\gamma^{*}\simeq3$.
This finding agrees with the boundary condition that $\gamma_{\kappa,T}\rightarrow3$
for $T\rightarrow+\infty$ and $\forall\kappa\in[0,1]$ (as stated
above, our model degenerates into the Barab\'{a}si-Albert model). Instead,
in the low temperature phase $T\ll T_{c}^{\kappa}$, any numerical
computation of the fugacity $z_{\kappa}$ suffers from finite size-effects,
but theoretical arguments put forward in \citep{Godreche2010} lead
to an estimate of $z_{\kappa}\simeq1$ also for the nonextensive case,
since $e_{\kappa}^{x}\sim e^{x}$ when $x\ll1$. Given that the average
fitness $\mathbb{E}(\eta)$ monotonically decreases to $0$ when the
temperature decreases, then from Eq. (\ref{eq:4}) the scaling exponent
$\gamma_{\kappa,T}$ has to increase when $T$ decreases, depending
on the nonextensive parameter $\kappa$. Nonetheless, as the \emph{lower}
bound in (\ref{eq:4b}) is valid only for degrees $k>k_{0}$, only
a part of the points on the $P(k)$ plot is shifted and raised from
the original behavior, at high degrees, and the result is a stretched
exponential, compatible with those shown in \ref{fig:Left:-Complementary-cumulative}. 

Furthermore, as reported in \citep{Godreche2010}, an accurate theoretical
analysis of the network dynamics in the condensate phase for the original
BB model $g(\epsilon)=2\epsilon,\,\epsilon\in(0,1)$ found evidence
for not only one ``gel'' node \citep{Krapiv00}, (i.e. one unique
condensate) but for an infinite hierarchy of condensates, whose degrees
grow faster than $F(\eta)$ and linearly with time. A similar picture,
in which there is not one unique ``winner'', was also presented
in \citep{Ferretti2008}.

This phenomenon is due to the presence of a crossover timescale $\tau(T)\sim T^{-2}$
in the condensate phase. The ``winner-takes-all'' phenomenon occurs
only for networks having evolved for times greater than $\tau(T)$,
which diverges at low temperature. For shorter times, a record-driven
dynamics \citep{Ferretti2008} is observed instead, with the number
$D(t)$ of candidate ``gel'' nodes up to time $t$ growing as $D(t)\sim2\ln\ln t$
\citep{Godreche2010}. These super-hubs, which sum up to the condensed
fraction $1-I_{\beta}$, escape our mean-field approach, in which
$k_{i}$ has to grow sublinearly in time, according to Eq. (\ref{eq:dynexp}).
Ultimately, it is those hubs which give rise to the stretched exponential
encountered in our numerical experiments in the condensate phase.

\section{Conclusions}

In this paper we have generalized the Barab\'{a}si-Bianconi fitness model
by the means of the non-Gaussian Kaniadakis $\kappa$-distribution,
which was originally proposed in the framework of nonextensive statistical
mechanics. Our analytical results show that the resulting generalized
fitness model presents a phase transition from a ``fit-get-rich''
phase to a ``gel'' phase, formally equivalent to a Bose-Einstein
condensation. Analytical calculations supported by numerical estimates
show that the critical temperature $T_{c}$ of the Bose-Einstein condensation
on networks decreases when the nonextensive parameter $\kappa$ is
increased from $0$ to $1$. A numerical analysis of the degree distribution,
complemented by an analytically obtained lower bound, reveals the
presence of power-law behaviour in the phase of high-temperatures
and a linear energy level density for $g(\epsilon)\rightarrow0$.
In contrast, in the condensate phase stretched exponentials, compatible
with the recent finding of a hierarchy of hubs, are retrieved.

\section{Acknowledgements}

MS is personally indebted to Prof. P. Tempesta and Prof. R. A. Leo
for the precious insights they provided. Without the latter, this
article would not have been written. MS is also grateful to Cl\'{e}ment
Viguier and the whole Complex Systems Simulation DTC at the Institute
of Complex Systems Simulation (ICSS), University of Southampton, for
their support. MS and MB also acknowledge the use of both MATLAB code
written by Aaron Clauset, from \citep{Clauset}, and of the IRIDIS
High Performance Computing Facility, and associated support services
at the University of Southampton, in the completion of this work.

\bibliography{RevisedFinale}
\bibliographystyle{ieeetr}

\end{document}